\documentclass[%
 reprint, superscriptaddress, nobibnotes,
 amsmath,amssymb, aps, pra
]{revtex4-2}

\usepackage{graphicx}
\usepackage{bm}
\usepackage{float}
\usepackage{hyperref}
\hypersetup{hidelinks}
\hypersetup{colorlinks = true, linkcolor=blue, urlcolor=blue,citecolor=blue}
\usepackage{physics}
\usepackage{suffix}
\usepackage{xstring}
\usepackage{siunitx}
\usepackage[normalem]{ulem}
\usepackage[dvipsnames]{xcolor}
\begin{document}
    
    \preprint{APS/123-QED}
    \title{Universal non-Gaussian order parameter statistics in 2D superfluids}
    
    \author{Abel Beregi}
    \email{abel.beregi@physics.ox.ac.uk}
    \affiliation{Clarendon Laboratory, University of Oxford, Oxford OX1 3PU, United Kingdom}
    
    \author{En Chang}%
    \affiliation{Clarendon Laboratory, University of Oxford, Oxford OX1 3PU, United Kingdom}
    
    \author{Erik Rydow}%
    \affiliation{Clarendon Laboratory, University of Oxford, Oxford OX1 3PU, United Kingdom}

    \author{Christopher J. Foot}%
    \affiliation{Clarendon Laboratory, University of Oxford, Oxford OX1 3PU, United Kingdom}
    
    \author{Shinichi Sunami}%
    \email{shinichi.sunami@physics.ox.ac.uk}
    \affiliation{Clarendon Laboratory, University of Oxford, Oxford OX1 3PU, United Kingdom}

\date{\today}

\begin{abstract}
Fluctuations are an intrinsic feature of many-body systems, and their full statistical distributions reveal a wealth of information about the underlying physics.
Of particular interest are non-Gaussian, extreme-value statistics that arise when nontrivial correlations and criticality dominate over the central  limit theorem.
Strikingly, in two-dimensional (2D) quantum fluids, such effects have been predicted to manifest in the order parameter distribution in the Berezinskii–Kosterlitz–Thouless (BKT) superfluid phase, which approaches a universal extreme-value form in the low-temperature limit.
Here, we measure the order parameter statistics of 2D Bose gases across the BKT critical point using matter-wave interferometry.
This allows us to confirm the predicted convergence of the observed statistics to a universal Gumbel distribution at low temperatures, to the 0.1\% level of the probability density.
Furthermore, the intrinsic precision of the atom interferometer allows the robust extraction of higher-moment observables such as skewness and kurtosis; in particular, we report direct measurements of the Binder cumulant which allows us to precisely identify the onset of the phase transition.
Extending this approach to the investigation of non-equilibrium systems, we probe vortex unbinding dynamics following a quench across the BKT critical point and identify parameter-independent scaling behaviour of higher moments. 
\end{abstract}

\maketitle

\section{Introduction}

The statistical distributions of fluctuating observables  provide a powerful means of characterising many-body systems, revealing emergent structures that cannot be inferred from average values alone~\cite{Sornette_2006}.
In many cases, Gaussian statistics arise as a consequence of the central limit theorem; yet strongly correlated and critical systems often display significant deviations from this paradigm, characterised by non-Gaussian, extreme-value distributions~\cite{Bramwell1998, bramwell_universal_2000}.
Such distributions have been reported across remarkably diverse areas of complex systems, ranging from Brownian motion~\cite{Farago_2002} to $1/f$ noise~\cite{antal_2001}, galaxies~\cite{antal_galaxy_2009}, atmospheric energy transfer~\cite{blender_fluctuation_2018}, protein sequences~\cite{webber_estimation_2001}, and evolutionary biology~\cite{kassen_distribution_2006}.
These widespread occurrences raise the fundamental question of when and how universal non-Gaussian distributions emerge from microscopic dynamics.

\begin{figure*}[t]
    \includegraphics[width=0.95	\linewidth]{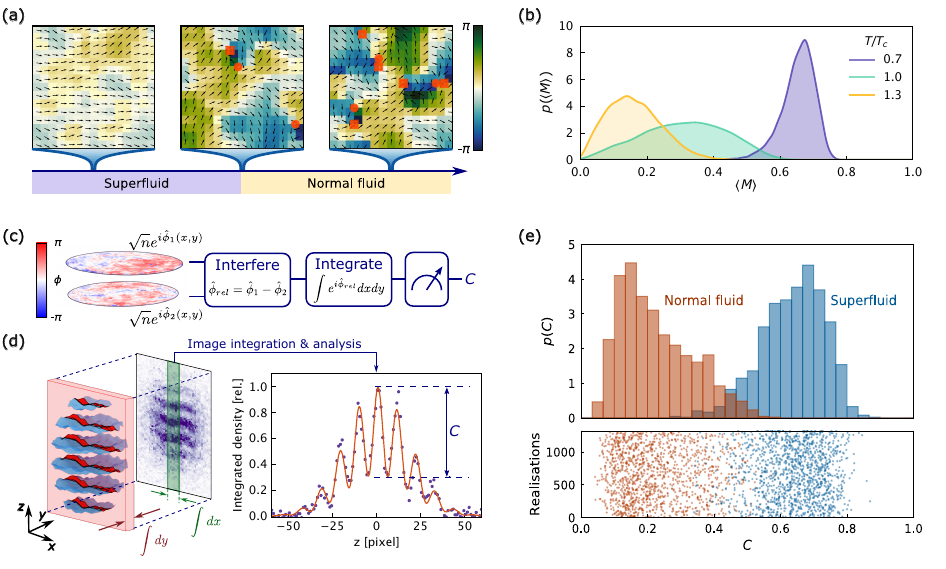}
    \caption{\label{fig:overview} 
    Matter-wave interferometric probe of order parameter statistics in the 2D XY universality class.
    (a), 2D XY model spin configurations at various temperatures across the BKT transition.
    At low temperatures ($T/T_c =0.7$), the system is quasi-ordered with only spin-wave excitations. 
    In the crossover region ($T/T_c =1.0$), vortex-antivortex pairs appear together with some unpaired vortices. 
    In the normal phase ($T/T_c =1.3$), vortex excitations proliferate (red circles and squares, representing vortices and antivortices). 
    (b), Numerically simulated order parameter distributions of the 2D XY model across the BKT critical point, exhibiting significant non-Gaussian features.
    The distribution in the superfluid regime has a universal Gumbel form.
    (c), Experimental protocol. 
    Two independent but identical 2D Bose gases are prepared, whose phase fluctuations are illustrated with the colour map.
    Matter-wave interferometry turns relative-phase fluctuations into interference patterns, which are integrated over a 2D area.
    This gives a single realization of the interferometric contrast $C$, which is sampled over many experimental realizations to obtain its full distribution.
    (d), Experimental implementation. After time-of-flight expansion of the 2D Bose gases, interference patterns are detected by absorption imaging;
    integration along $y$ is realised intrinsically by absorption imaging with a controlled depth (red sheet indicates selective repumping~\cite{Sunami2022}), and along $x$ by image post-processing. 
    The contrast is extracted from the integrated interference signal.
    (e), Examples of observed contrast statistics across the BKT transition, exhibiting non-Gaussian features. 
    More than 1000 samples are obtained from over 200 experimental repeats. 
    }
\end{figure*}

A crucial paradigm is provided by condensed-matter systems, in which full-distribution functions provide key insights into the connection between fluctuations and nontrivial collective behaviour, independent of the microscopic details of the systems~\cite{Polkovnikov2006, Rath2010, ivanov_characterizing_2013}.
In particular, the full distribution function (FDF) of the degree of coherence, quantified by an order parameter such as the global magnetisation or macroscopic wavefunction, is of key importance for such investigations~\cite{archambault_universal_1998, lamacraft_order_2008, bramwell_distribution_2009}.
Systems in which phase transitions arise from spontaneous symmetry breaking are expected to exhibit non-Gaussian FDFs of the order parameter near critical points, where fluctuations dominate and harmonic approximations fail.
In low-dimensional systems, the effect of fluctuations is significant at any non-zero temperature~\cite{Mermin1966,Hohenberg1967}, leading to rich emergent behaviour.
The two-dimensional (2D) XY universality class is an archetype of universality in low-dimensional systems associated with the Berezinskii–Kosterlitz–Thouless (BKT) transition: a vortex-mediated topological phase transition between a low-temperature superfluid phase with quasi-long-range order and a disordered phase at high temperatures~\cite{Berezinskii1972,Kosterlitz1973,Jose2013} (Fig.~\ref{fig:overview}a). 
As a result of scale-invariant fluctuations in the BKT superfluid phase, the order parameter FDF is expected to be a universal extreme-value distribution~\cite{bramwell_analytic_2022,Rath2010} (Fig.~\ref{fig:overview}b).

Despite the fundamental importance of order parameter FDFs, their direct experimental access has been challenging, especially in conventional solid-state systems \cite{Lovas2017}.
Ultracold atomic gases provide a versatile platform for FDF measurements of many-body quantum systems, with high controllability of system parameters and geometry, as well as reliable preparation of many-body states to obtain large statistics for identifying the underlying distribution~\cite{Schafer2020}.
The development of measurement techniques for probing order parameter FDFs in ultracold atomic systems is an active field of research, with recent advances in single-atom-resolved full-counting approaches in mesoscopic systems~\cite{xiang_situ_2025, yao_measuring_2025, de_jongh_quantum_2025} and optical-lattice experiments~\cite{herce_full_2023,wienand_emergence_2024,impertro_local_2024,allemand2025}.
In contrast, for larger systems with tens of thousands of particles comprising a continuous quantum field, matter-wave interferometry (MWI) is a powerful approach that provides direct access to microscopic fluctuations~\cite{Shin2004,Hadzibabic2006,Polkovnikov2006,Imambekov2008,Hofferberth2008,beregi_2024,Sunami2022,rydow2025}.
Major advantages of the atom interferometry approach are the intrinsic precision and robustness against experimental noise, often exploited for quantum metrology~\cite{cronin2009,Dimopoulos2009,Overstreet2022}; these allow full-distribution measurements with low noise levels and large statistics, and enabled the contrast FDF approach for probing one-dimensional quantum gases~\cite{Hofferberth2008,Kuhnert2013multimode,Gring2012}.
In 2D, the interference contrast gives access to the order parameter FDF across the BKT transition~\cite{Polkovnikov2006, Rath2010}, connecting the universal FDF and interferometry readout~\cite{Rath2010,bramwell_analytic_2022}.
Therefore, the MWI-based FDF measurements of 2D continuous systems represent an exciting prospect not only for probing the equilibrium universal distribution but also for uncovering higher-order effects in non-equilibrium dynamics and their possible scaling behaviour.

In this work, we report measurements of matter-wave interferometric contrast distributions of decoupled bilayer quantum gases, which allow us to probe directly the order parameter FDF of 2D Bose gases across the BKT critical point (Fig.~\ref{fig:overview}c-e).
We first observe the contrast distribution, which shows distinct modifications in its shape across the BKT critical point.
At low temperatures, we identify a robust universal extreme-value distribution of the order parameter in accordance with theoretical predictions~\cite{bramwell_universal_2000}.
The FDF measurement further allows the extraction of higher moments such as the skewness and kurtosis of the order parameter, and we demonstrate the direct observation of the Binder cumulant with characteristic crossover behaviour consistent with numerical simulations.
Next, we apply this robust contrast distribution approach in a non-equilibrium setting, enabled by the controllability of our bilayer trap.
In particular, we study the FDF of bilayer 2D Bose gases following a coherent splitting, which rapidly quenches the system from the superfluid to the normal phase across the BKT critical point, as recently investigated via the two-point correlation function approach to reveal parameter-independent universal dynamics described by real-time renormalisation group theory~\cite{Sunami2023}.
The FDF probe reveals the parameter-independent scaling dynamics of higher-order correlations, confirmed by the collapse of the time evolution of higher moments of the contrast FDF.
Finally, we identify the non-thermal nature of these distributions by direct comparison with equilibrium states.

\section{Contrast full distribution measurements of 2D quantum gases}
\label{sec:equilibrium}

To probe the statistical distribution of the fluctuating 2D quantum fields, we prepare bilayer 2D degenerate Bose gases of ${}^{87}$Rb atoms in the lower hyperfine level in a cylindrically-symmetric double-well trap, created by the multiple radio-frequency (MRF) dressing technique \cite{Sunami2022, Sunami2023, rydow2025}.
2D Bose gases are in the 2D XY universality class~\cite{Posazhennikova2006,Hadzibabic2006}, and detailed experimental observation of the BKT transition has been made in recent years through the direct probe of the two-point correlation functions and vortex statistics in our setup~\cite{Sunami2022, rydow2025}.
Strong axial confinement along the $z$-direction in each of the two wells ensures the quasi-2D condition for the trapped gases.
The axial trapping potential is harmonic near the bottom of the two wells, with a vertical trap frequency of $\omega_z/2\pi \approx \SI{1.1}{\kilo\hertz}$ and the separation between them being $\Delta z \approx \SI{4.9}{\micro\meter}$, realising a pair of independently fluctuating 2D quantum gases~\cite{Sunami2022}. 
The dimensionless 2D interaction strength is $\tilde{g} = \sqrt{8\pi} a_s/\ell_0=0.09$, where $a_s$ is the 3D s-wave scattering length, $\hbar$ is the reduced Planck constant, and $\ell_0=\sqrt{\hbar/(m\omega_z)}$ is the harmonic oscillator length along $z$ for an atom of mass $m$.
A box-shaped optical trap in the $x$-$y$ plane, created using a strong off-resonant laser beam with a ring-shaped profile, generates a near-homogeneous 2D system~\cite{Sunami2024}.
The cloud is cooled by forced RF evaporation to a temperature $T=\SI{32}{\nano \kelvin}$; by varying the atom number $N$ in each layer, which is made to be balanced by maximising the interference contrast~\cite{Barker2020}, we change the phase-space density $\mathcal{D} = n h^2 /(2\pi m k_B T)$ over the range $5$ to $21$, where $n$ is the average 2D density and $k_B$ is the Boltzmann constant.
In all cases, the quasi-2D conditions $\hbar \omega_z > k_B T$ and $\hbar \omega_z > \mu$ are satisfied, where $\mu$ is the chemical potential. 

To probe the system, the trap is turned off abruptly, releasing the pair of 2D gases for a time-of-flight (TOF) expansion of duration $t_{\text{TOF}}=\SI{17}{\milli \second}$.
The released clouds expand rapidly along the $z$-axis and overlap, forming an interference pattern along $z$ (Fig.~\ref{fig:overview}d), whose spatial phase profile in the $x$-$y$ plane encodes fluctuations of the relative phase mode of the in-situ system \cite{Pethick2008,Sunami2022}. 
Before absorption imaging, a laser light sheet of thickness $L_y = \SI{8}{\micro\metre}$ optically pumps atoms from the lower to the upper hyperfine level in a slice indicated by the red transparent sheet in Fig.~\ref{fig:overview}d, which are then detected~\cite{Barker2020,Sunami2022}.
\begin{figure*}[t]
    \includegraphics[width=\linewidth]{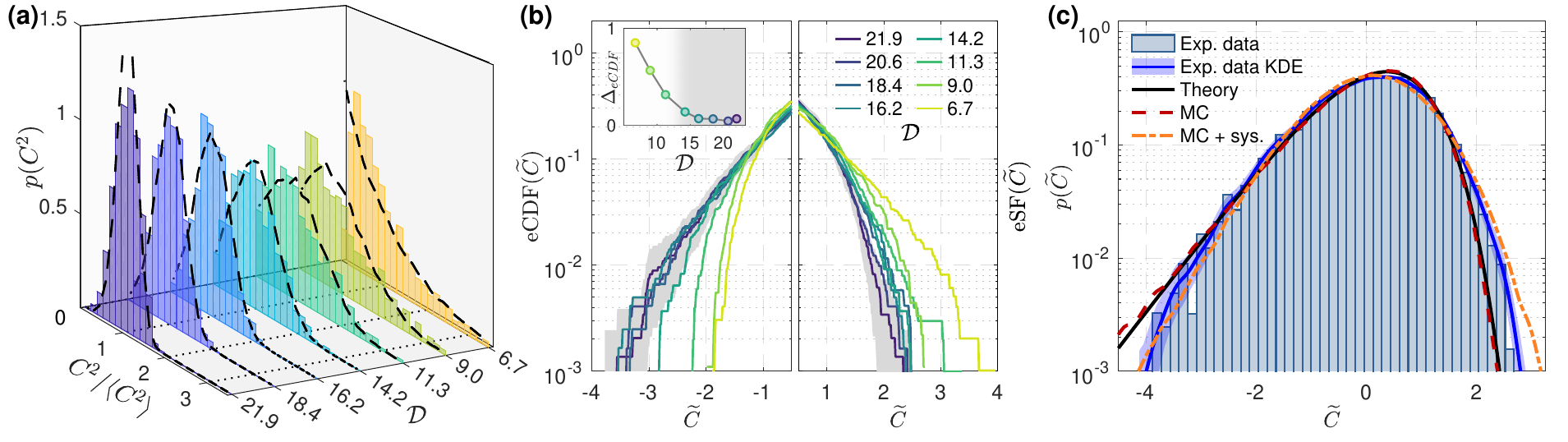}    \caption{\label{fig:equilibrium_bkt_contrast} 
    Convergence of order parameter statistics to universal Gumbel distribution.
    (a), Probability histograms of the experimental $C^2$ at various values of $\mathcal{D}$ compared to Monte Carlo (MC) simulations (dashed lines). 
    (b), Empirical cumulative distributions (eCDF) and survival functions (eSF), each obtained from at least 900 contrast values, plotted on a logarithmic scale to highlight the collapse of the distributions onto a universal curve over several orders of magnitude.
    The shaded area indicates the 99 percent confidence interval for the highest $\mathcal{D}$ dataset. 
    Inset shows the scaled sum-of-squared residuals between the eCDF of the low temperature MC simulation result and each experimental dataset. The shaded region indicates the identified universal regime. 
    (c), Universal order parameter statistics in the low-temperature limit. Histogram of the experimental data within the universal regime ($14 \leq\mathcal{D} \leq 21$), containing over 7500 contrast values. The solid blue line is a kernel density estimate (KDE) and the shaded area indicates its 99 percent confidence interval, obtained via bootstrapping. The dashed lines are from Monte Carlo simulations for $14 \leq\mathcal{D} \leq 21$, analysed with (orange) and without (red) accounting for the systematic effect of imaging noise (Methods).
    }
\end{figure*}
From the ensemble of acquired images, we calculate the statistical distribution of the spatially integrated interferometric contrast.
Integration along $y$ arises intrinsically in absorption imaging that integrates the density along the imaging axis and along $x$ through post-processing of the images.
The obtained data are the integrated density profiles along $z$, which can be analysed by their Fourier spectra (Methods).
The contrast value associated with each density profile represents the visibility of the interference fringes, as illustrated in Fig.~\ref{fig:overview}d.
The high-momentum region of the spectrum further allows for the reconstruction of intrinsic noise statistics of the imaging process for a comprehensive analysis of the systematic effects of our protocol (see Supplementary Information).

In Fig.~\ref{fig:overview}e, we show examples of experimental contrast statistics for 2D Bose gases in the superfluid and normal regimes, displaying distinct non-Gaussian characteristics.
In the superfluid regime, the presence of quasi-long-range order results in interference fringes with high visibility, whose distribution shows a long tail towards low contrast.
Conversely, in the normal regime, the strong phase fluctuations reduce the mean observed contrast and increase the width of the distribution.
There is a finite chance of finding vortex-free regions with partial coherence in the system; thus, the observed distribution has a long tail towards higher contrasts. 
In the following, we perform an in-depth analysis of the full distributions of measured contrast, which gives quantitative detail about the previously inaccessible higher-moment observables of the 2D quantum gases.

\section{Universal order parameter distribution of BKT superfluids}
\label{sec:universal}

The squared contrast observable associated with the interference of two 2D bosonic fields is given by~\cite{Polkovnikov2006}
\begin{equation}
    	\hat{C}^2 = \frac{1}{N_1 N_2} \int_\Omega d\mathbf{r}_1 d\mathbf{r}_2 \hat{\psi}_1^\dagger(\mathbf{r}_1)\hat{\psi}_1(\mathbf{r}_2)\hat{\psi}_2(\mathbf{r}_1) \hat{\psi}_2^\dagger(\mathbf{r}_2), 
        \label{eq:C_squared observable}
\end{equation}
where $\hat{\psi}_{i}^{(\dagger)}$ are the bosonic annihilation (creation) operators in layer $i=1,2$, and $N_i$ is the number of particles; the region of integration $\Omega$ is determined by the thickness of the repumping light sheet and the integration width used for image post-processing.
For two independent yet identical Bose gases in the quasicondensate regime, covering both normal and superfluid phases of the BKT transition~\cite{Prokofev2001}, Eq.~\eqref{eq:C_squared observable} simplifies to
\begin{equation}
    	  C  = \frac{1}{N} \int_\Omega d\mathbf{r}~ \bar{n}(\mathbf{r})e^{i(\phi_1(\mathbf{r})-\phi_2(\mathbf{r}))},
        \label{eq:C_observable}
\end{equation}
where, within the classical field approximation, we replaced the field operators with complex numbers in the density-phase representation.
Since density fluctuations are energetically suppressed in the quasicondensate regime, the mean density $\bar{n}(\mathbf{r})$ is introduced.
Within these approximations, the interference contrast depends only on the relative phase mode which, for uncoupled layers, has the same statistics as the phases of the individual layers $\phi_i(\mathbf{r})$, but with twice the magnitude~\cite{Sunami2022}.
Thus, $C$ in Eq.~\eqref{eq:C_observable} is equivalent to the order parameter for a single layer 2D Bose gas, $A = \int d\mathbf{r} \psi(\mathbf{r})$, and differs only by a prefactor of 2; this quantity corresponds to the magnetisation in the 2D XY model, thereby establishing a clear link between interference contrast statistics and order parameter statistics. 

In Figure~\ref{fig:equilibrium_bkt_contrast}a, we show the observed FDF of the squared contrast for a wide range of phase-space densities across the BKT critical point~\cite{Sunami2022,rydow2025}.
Contrast values are normalised by their root-mean-squared value for each dataset, thus fixing the mean of $C^2$ to unity, highlighting the change in the shape of the distributions.  
The distributions are compared with classical-field Monte Carlo simulations of interference contrast, which account for the well-characterised technical noise affecting the contrast extraction procedure (see Supplementary Information).
We observe good agreement between the experimental and simulated distributions throughout the BKT crossover.
Deep in the superfluid regime, where $\mathcal{D}$ is significantly above the critical value $\mathcal{D}_c = 11.5(7)$, the probability density $p(C^2)$ exhibits a sharp peak centred around unity, which is skewed toward low contrasts. 
In the normal phase, the probability of observing vanishing contrasts is the largest, and the distribution is Poissonian.
Theoretical works predict such a change in $p(C^2)$~\cite{Polkovnikov2006,Imambekov2008,Rath2010}, in good agreement with our experimental observation.

In the superfluid regime, the FDF of the contrast observable is expected to have a universal functional form given by the Gumbel distribution, upon normalisation of the mean and variance, which are non-universal~\cite{Rath2010}.
To see this, we introduce the variable $\widetilde{C} = (C-\langle C \rangle)/\sqrt{\langle C^2 \rangle}$, which makes the statistical analysis more sensitive to the shape of the distributions arising from third- and higher-order moments. 
To eliminate the effect of binning, we compute the empirical cumulative distribution function (eCDF) as well as the empirical survival function (eSF = 1-eCDF) for each dataset (see Supplementary Information). 
These are shown in Fig.~\ref{fig:equilibrium_bkt_contrast}b: eCDFs sensitively characterise the left tail of $p(\widetilde{C})$ and eSFs for the right tail of $p(\widetilde{C})$, when plotted on a logarithmic scale. 
The data demonstrate a collapse towards a universal distribution in the superfluid regime, within the 99 percent confidence intervals indicated by the grey shaded areas, down to the 0.1\% level (see Supplementary Information). 
To quantify the collapse, we calculate the sum of squared residuals between the eCDF obtained from the Monte Carlo simulation results in a regime where universal behaviour was present, and each experimental dataset, denoted as $\Delta_{eCDF}$ (inset).
The obtained values of $\Delta_{eCDF}$ remain small and constant deep in the superfluid regime, indicating the robust collapse of the distributions to a common one for high PSDs.
Furthermore, this shows that the experimental distributions in this regime agree well with the low temperature distribution from the Monte Carlo simulations.
We observe that universality holds over a range of $\mathcal{D} \gtrsim 14.0$, in good agreement with theoretical predictions for 2D Bose gases \cite{Rath2010}.
Since the mean and variance are normalised by the use of $\widetilde{C}$, the collapse of the distributions arises solely from the universal behaviour of all higher moments, providing evidence for the universality of 2D Bose gases in the low-temperature regime beyond two-point correlation functions.

In Fig.~\ref{fig:equilibrium_bkt_contrast}c, we show the combined statistics of the experimental contrast obtained in the universal regime ($\mathcal{D} \geq 14.3$), totalling over 7500 samples, along with the corresponding kernel density estimate (KDE) and its confidence intervals (see Supplementary Information). 
The large sample size, enabled by the high repeatability of our cold-atom experiment, allows for extensive comparisons with the theoretical model.
This universal probability density function corresponds to the low-temperature order parameter distribution of the 2D XY model in thermal equilibrium, and as such has been studied extensively \cite{archambault_universal_1998,bramwell_magnetic_2001}.
Analytical results in the thermodynamic limit at low temperatures show a Gumbel extreme value distribution with known parameters~\cite{bramwell_universal_2000, bramwell_analytic_2022}, along with small corrections for finite-size systems~\cite{banks_temperature-dependent_2005,palma_finite-size_2016}.
In Fig.~\ref{fig:equilibrium_bkt_contrast}c, we also show this analytical result (black solid curve).
In the standard language for the 2D XY model, the universal low-temperature distribution has the form
\begin{equation}
    p(\tilde{m}) = K(e^{{x}-e^x})^a, \ \ \ x=b(\tilde{m}-s),
    \label{eq:bhp_dist}
\end{equation}
where $\tilde{m} = (m-\langle m \rangle )/ \sqrt{\langle m^2 \rangle }$ is the normalised magnetisation equivalent to $\widetilde{C}$, and $K=2.14$, $a=\pi/2$, $b=0.938$, $s=0.374$ are universal constants~\cite{bramwell_universal_2000}.
Further, we show the Monte Carlo simulation results (red dashed curve), which are consistent with the analytical prediction.
After incorporating the systematic effect of the well-characterised imaging noise into the Monte Carlo results (Supplementary Information), we observe good agreement between the simulation and the experiment, showing the consistency of the analytical Gumbel-type distribution with the 2D XY model and the experimental distribution.

Finally, the distribution shown in Fig.~\ref{fig:equilibrium_bkt_contrast}c enables a quantitative analysis of the two relevant moments of the observed universal distribution to be compared with theoretical predictions for the 2D XY model~\cite{banks_temperature-dependent_2005,palma_temperature_2006}.
The skewness of the experimental data $\gamma_1^{exp}=-0.55-0.15(sys.)\pm0.07$ is in agreement with the finite-temperature Monte Carlo estimate $\gamma_1^{MC}=-0.76$, after accounting for the systematic effect of imaging noise \cite{banks_temperature-dependent_2005}.
Similarly, we calculate from the experimental data that the kurtosis is $\gamma_2^{exp} = 3.1+0.6(sys.)\pm0.3$ in agreement with its predicted finite-temperature value $\gamma_2^{MC}=3.99$ \cite{palma_temperature_2006}.

\begin{figure}[t]
    \includegraphics[width=0.99	\linewidth]{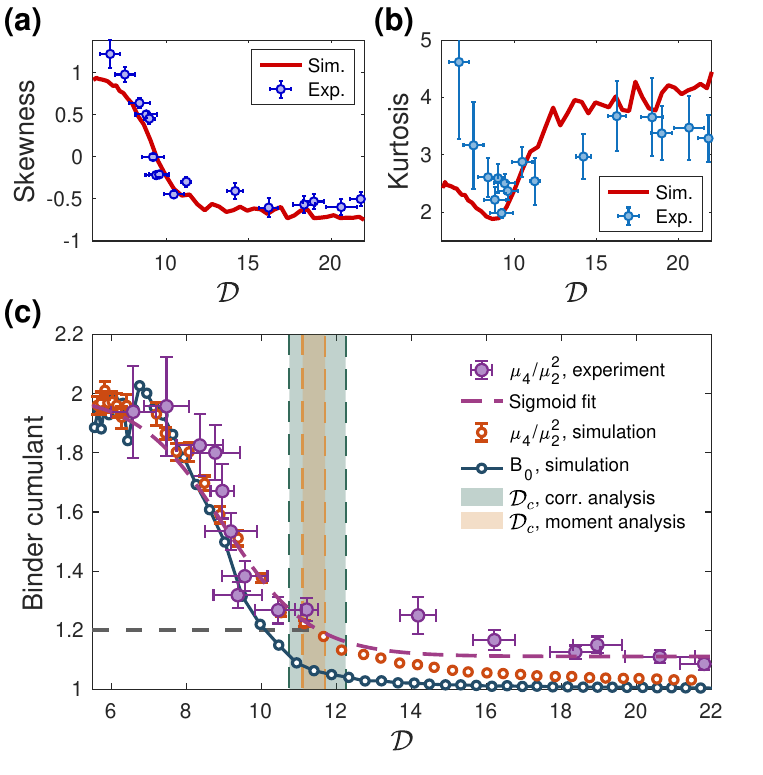}
    \caption{\label{fig:equilibrium_binder} 
    Measurements of higher-moment observables associated with contrast FDFs across the BKT critical point. 
    (a),(b), Skewness and kurtosis of the normalised distribution $p(\widetilde{C})$. 
    The simulation data (red solid lines) is from moments obtained by Monte Carlo simulations of interference contrast statistics.
    (c), Binder cumulant measurement from the raw moments of full distribution functions of interference contrast. 
    Experimental data (purple markers) and simulated data (red open markers) exhibit the characteristic step-like behaviour from 1 to 2 across the BKT critical point.
    The blue open markers show the result from direct integration of the individual fields from Monte Carlo simulations (the connecting lines are guide to the eye).
    The green shaded area marks the experimentally detected critical phase-space density from the complementary analysis of the two-point phase correlation functions. 
    The yellow shaded area indicates the experimentally detected critical phase-space density from the analysis of Binder cumulants.
    The horizontal dashed line shows the critical Binder cumulant, as predicted from the finite-size scaling analysis (Supplementary Information).
    }
\end{figure}

\section{Higher-moment observables across the critical point}
\label{sec:higher}
Until now, our analysis has focused on the universality of the contrast FDF deep in the superfluid regime. 
However, the distribution $p(\widetilde{C})$ provides especially valuable information about the state of the system in the crossover region.
In Fig.~\ref{fig:equilibrium_binder}a-b, we analyse the evolution of $p(\widetilde{C})$ across the BKT critical point by examining its first two relevant moments, the skewness and the kurtosis. 
Deep in the superfluid regime, the values of these higher-moment observables remain constant within experimental error, consistent with the universality of the FDFs investigated earlier.
As the system approaches the critical point from high $\mathcal{D}$ towards the normal phase, the skewness increases sharply and undergoes a sign reversal, from negative to positive.
The kurtosis also exhibits characteristic behaviour near the critical point, reaching a pronounced minimum.
Remarkably, the inversion of skewness and the minimum of kurtosis occur at the same PSD near the critical point, where the distribution is closest to a Gaussian.
This demonstrates a fundamental distinction between the topological BKT phase transition and spontaneous symmetry breaking, as in the latter case, the order parameter distribution is Gaussian far from the critical point.

\begin{figure*}[t]
    \includegraphics[width=0.75	\linewidth]{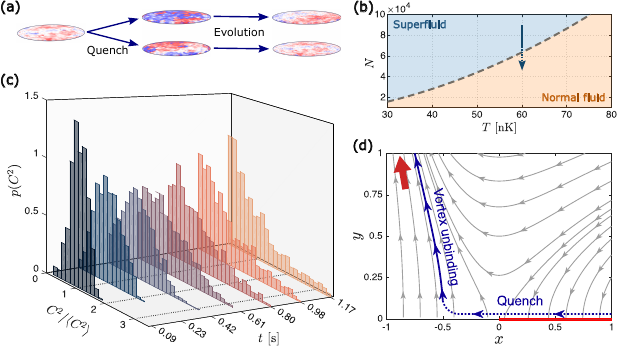}
    \caption{\label{fig:quench} 
     Probing out-of-equilibrium dynamics of 2D Bose gases with interference contrast full distribution measurements.
     (a), Schematic of the experimental protocol. A single layer 2D Bose gas is split coherently into two clouds. The decoupled gases are then evolved for variable amount of time before probing the system with a time-of-flight interferometer. 
     (b), The coherent splitting quench shown on the phase diagram.
     (c), Probability histograms of the experimental $C^2$ at various times after the quench. 
     (d), Real-time renormalization group flow diagram for the 2D XY model. 
     The red arrow and the red line (for $x>0,\ y=0$) indicate the fixed points of the flow. 
     The quench induces the equilibrium states in the superfluid phase to be driven into the normal phase (dotted blue trajectory), followed by dynamical vortex unbinding (solid blue trajectory).
    }
\end{figure*}

In addition to the observation of the universal distribution of the normalised contrast $\widetilde{C}$ and its crossover behaviour, the analysis of the \textit{unnormalized} distribution provides more direct information on the statistics of the order parameter $A$ through Eq.~\eqref{eq:C_observable}. 
The extraction of raw moments allows us to identify the Binder cumulant $B_0$~\cite{Binder1981} 
\begin{equation}
    B_0 = \frac{\langle \vert A \vert ^4 \rangle}{\langle \vert A \vert^2 \rangle^2} \approx \frac{\mu_4}{\mu_2^2}, 
    \label{eq:binder_vs_moments}
\end{equation}
where $\mu_n$ is the $n$-th raw moment of the contrast distribution $p(C)$.
$B_0$ is a quantity of central importance for the critical behaviour of many-body systems: for the 2D XY model, in the thermodynamic limit, it exhibits a jump from 1 to 2 at the BKT phase transition.
For finite-sized systems, $B_0$ shows a crossover from 1 to 2, and its critical value shows small logarithmic corrections at the $10^{-2}$ level \cite{loison_binders_1999, hasenbusch_binder_2008}.
This parameter has been used to characterise the critical region and precisely identify the critical point in many numerical works relating to 2D and 3D Bose gases in and out of equilibrium~\cite{keepfer2022, comaron_quench_2019, kobayashi_thermal_2016, kobayashi_quench_2016, liu_kibble-zurek_2020}.
Although the Binder cumulant is of broad theoretical importance, its experimental determination has been scarce, motivating ongoing efforts such as recent work on heavy-ion collisions~\cite{sorensen_locating_2024}. 
The well-controlled sample preparation and matter-wave interferometric readout of the order parameter achievable in ultracold atomic systems provide a robust route for reliably accessing this quantity.

In Fig.~\ref{fig:equilibrium_binder}c, we plot the Binder cumulant extracted from the experimental contrast distributions as the $\mu_4/\mu_2^2$ ratio of raw moments, exhibiting the characteristic crossover from 1 to 2 across the BKT critical point (purple filled markers).
The experimental data show good agreement with the same moment ratio from Monte Carlo simulations of interference contrast statistics in the crossover regime (red open markers).
The simulated fields are also used to evaluate the reference values for the Binder cumulant, $B_0$, computed directly from the statistics of the order parameter (blue connected open markers).
In the superfluid regime, the cumulant extracted from noisy contrast statistics yields a marginally higher value than $B_0$; this is due to the finite systematic effect of photon shot noise in the imaging process, which increases $\mu_4$ faster than $\mu_2^2$.
Fitting the experimental data with a sigmoid function, we identify the critical phase-space density by evaluating the point where the fitted curve crosses the value of 1.2, the critical Binder cumulant predicted by our Monte Carlo simulations (see Supplementary Information) and other recent numerical results~\cite{keepfer2022}.
The obtained value $\mathcal{D}_c = 11.4(3)$ agrees well with the critical phase-space density from the complementary analysis of the two-point phase correlation function.
This highlights the usefulness of using the Binder cumulant experimentally for precision measurements of the critical point, with robustness against calibration, finite-size~\cite{loison_binders_1999} and finite sampling effects~\cite{kikuchi1996}, including wider applications such as 2D Bose gases in the presence of disorder~\cite{Carleo2013}.

\section{FDF measurements of non-equilibrium 2D systems}
\label{sec:quench}

\begin{figure*}[t]
    \includegraphics[width=0.9	\linewidth]{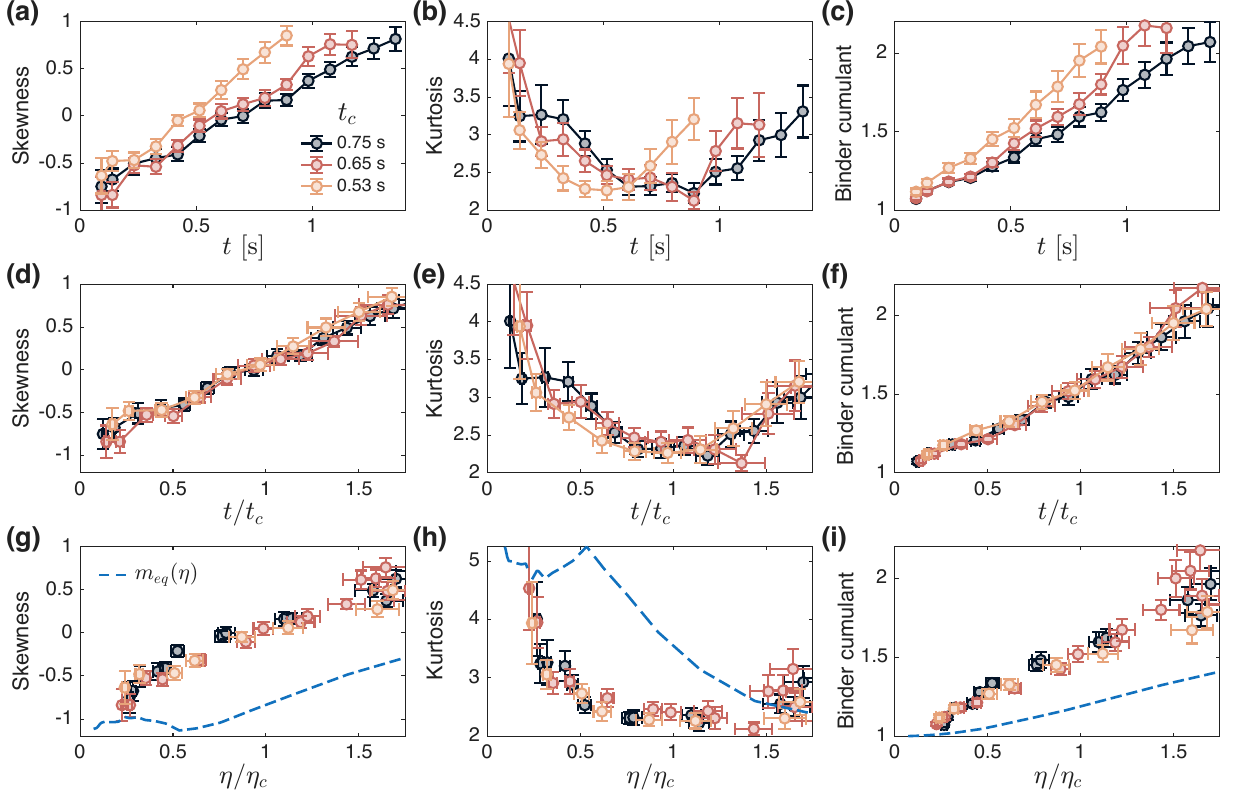}
    \caption{\label{fig:quench_moments} 
     Out-of-equilibrium dynamics of higher-moment observables corresponding to the order parameter of 2D Bose gases following a quench across the BKT transition. 
     (a)-(c), Temporal evolution of skewness, kurtosis and Binder cumulant as a function of time for three different initial conditions, indicated by crossover times $t_c$. 
     (d)-(f), Universal dynamics of higher-moment observables in reduced time $t/t_c$.
     (g)-(i), Comparison of universal dynamics of higher-moment observables with equilibrium simulations through the mapping $f: \eta \rightarrow m_{eq}$ (blue dashed line).
     }
\end{figure*}

So far, we have investigated the FDF of 2D Bose gases in equilibrium, identifying the universal extreme-value statistics of BKT superfluids and the crossover behaviour of higher moments across the critical point.
Out-of-equilibrium 2D Bose gases exhibit a wide range of novel phenomena, such as the relaxation of turbulence \cite{kwon_relaxation_2014, reeves_turbulent_2022, karailiev_observation_2024}, critical dynamics near a non-thermal fixed point \cite{schole_critical_2012, karl_strongly_2017, rasch_decaying_2025}, Kardar-Parisi-Zhang universality \cite{diessel_emergent_2022, deligiannis_kardar-parisi-zhang_2022, widmann_observation_2025}, coarsening dynamics \cite{comaron_quench_2019,Groszek2021, gazo_universal_2025, chang2025}, and dynamical vortex unbinding \cite{Mathey2017, Sunami2023}, to name a few. 
Full characterisation of such diverse non-equilibrium states remains a major challenge; 
the interferometry-based FDF measurement presented in this work, which extends to all integrated higher-order correlation functions~\cite{Polkovnikov2006}, is a very promising approach for uncovering such nontrivial dynamics~\cite{AduSmith2013, steffens_towards_2015, gopalakrishnan_distinct_2024}.

Here, as a representative non-equilibrium system, we investigate the order parameter FDF of 2D Bose gases following a quench across the BKT critical point, which creates highly non-equilibrium, superheated superfluid states that relax via a universal vortex unbinding process~\cite{Mathey2017,Sunami2023}.
The quench is implemented by a \emph{coherent splitting} procedure, in which a single 2D gas is split into two to reduce the density suddenly, thereby crossing the critical point from the superfluid to the normal phase for appropriately chosen initial conditions, as illustrated in Figure~\ref{fig:quench}a-b.

Figure~\ref{fig:quench}c shows the time-dependent FDFs of the squared contrast observable at various times following the quench.
Similar to the equilibrium data, the contrast values for each dataset were normalised to fix the mean of $C^2$ to unity. 
This shows a \textit{temporal} evolution of $p(C^2)$ from a peaked distribution to a Poissonian, with the same general features across the equilibrium BKT critical point, with decreasing PSDs (Figure~\ref{fig:equilibrium_bkt_contrast}a).
However, these distributions originate from highly non-equilibrium states and corresponding order parameter statistics, which are expected to be distinct from equilibrium states.
A useful theoretical framework for understanding the relaxation of low-order observables is the real-time renormalisation-group (RG) theory~\cite{Mathey2010, Mathey2017, Sunami2023} (Fig.~\ref{fig:quench}d). 
In this picture, the many-body dynamics is determined by the spatiotemporal renormalisation of the variables $x=1/(2\eta)-1/(2\eta_c)$ and $ y=\sqrt{2}\pi g$, where $\eta_c$ is the critical value of the power-law exponent $\eta$, and $g$ is the vortex fugacity.
The RG equations read as 
\begin{equation}
    \begin{split}
        & \frac{dx}{d t} = -\frac{(x+2)^3 y^2}{8t},   \\
        & \frac{dy}{d t} = -\frac{xy}{t}, 
    \end{split}
    \label{eq:real_time_rg_eqns}
\end{equation}
representing the universal evolution~\cite{Mathey2017,Sunami2023} (see Supplementary Information).
Within this treatment, possible equilibrium states are characterised by the fixed points of the flow, as indicated by the red arrow for the high-temperature phase and the red line for the low-temperature phase ($x>0$) in Fig.~\ref{fig:quench}d.
As experimentally observed in Ref.~\cite{Sunami2023} through the analysis of two-point correlation functions and vortex excitations, the coherent splitting quench and rapid prethermalization (dashed arrow) drive the system from the superfluid fixed points in $x>0$ to initiate non-equilibrium vortex unbinding dynamics (solid arrow), eventually reaching the normal phase with $y\gg 1$ in the long-time limit. 
The late-time relaxation of the system is consistent with spatiotemporal scaling, which is robust against the choice of initial state before the quench by the rescaling of time with the crossover time $t_c$, a characteristic timescale for the dynamics, the time at which the two-point correlation loses its quasi-long-range coherence~\cite{Sunami2023}.
Below, we extend the analysis to higher-moment observables to see whether the scaling behaviour persists for higher-order correlations.

In Fig.~\ref{fig:quench_moments}a-c, we show the time evolution of the higher-moment observables of the quenched systems with three different initial conditions (PSDs).
While all initial conditions result in the crossing of the critical point by the splitting quench, the strength of the quench differs, resulting in distinct crossover times $t_c = 0.53(4) ~\text{s}, \ 0.65(6)~\text{s, and} \ 0.75(4)~\text{s}$.
The temporal evolution of the skewness and kurtosis shows the same qualitative behaviour for all three initial conditions. 
Namely, the zero crossing of the skewness and the minimum of kurtosis occur near $t_c$, with faster dynamics for stronger quenches.
This feature is also shared by the Binder cumulant, which shows a smooth evolution from 1 to 2 across $t_c$.
Remarkably, the observed dynamics of the higher-moment observables collapse onto a single curve through the temporal rescaling $t \rightarrow t/t_c$ (Fig.~\ref{fig:quench_moments}d-f), connecting the initial conditions, time, and higher-order correlations via a scaling function akin to the scaling dynamics of the low-order two-point correlation function~\cite{Sunami2023}.

An interesting feature of these dynamics is that the observed scaling of the higher-moment observables cannot be inferred from lower-order correlation functions alone, demonstrating the non-equilibrium nature of the contrast FDF.
In equilibrium, i.e., at the fixed points of the RG flow (Fig.~\ref{fig:quench}d), different states in the superfluid and crossover regimes, characterised by different algebraic exponents $\eta$, have a one-to-one mapping to higher-moment observables (Fig.~\ref{fig:equilibrium_binder}), allowing for the construction of mappings $f: \eta \rightarrow m_{eq}, \ m_{eq}=\{\gamma_1, \gamma_2, B_0\}$, for the skewness, the kurtosis and Binder cumulant respectively.
If the non-equilibrium system had a corresponding but time-dependent order parameter FDF in equilibrium, then the measured observables $m(t)$ would coincide with the equilibrium values $m_{eq}$ after the mappings $f$ at all times.
However, as we show in Fig.~\ref{fig:quench_moments}g-i, the mapped $m_{eq}$ significantly differs from the observed dynamics of higher-moment observables, demonstrating universal yet highly non-trivial fluctuations during the relaxation dynamics.
In particular, the absence of a plateau in the skewness and kurtosis throughout the low-$\eta$ regime represents a qualitative change of the scaling behaviour between the equilibrium and out-of-equilibrium scenarios.
This feature is also shared by raw moments up to order four, as demonstrated using the Binder cumulant in Fig.~\ref{fig:quench_moments}i.

\section{Conclusion}
\label{sec:conclusion}

In conclusion, we have probed the full distribution of order parameter fluctuations in 2D quantum gases using matter-wave interferometry.
We confirmed the emergence of the universal extreme-value distribution deep in the superfluid regime, in excellent agreement with the predicted Gumbel distribution. 
Using the skewness and kurtosis of order parameter statistics, as well as the Binder cumulant, we characterised the BKT crossover beyond the two-point phase correlations and found good agreement with classical-field Monte Carlo simulations. 
Extending these methods to a non-equilibrium scenario, we probed the relaxation dynamics of 2D Bose gases following a quench from the BKT superfluid to the normal phase through measurements of the full distribution functions of interference contrast. 
The evolution of the higher moments in rescaled time shows a collapse onto parameter-independent behaviour, indicating the extension of a universal description to higher-order observables of dynamical processes, which calls for a corresponding theory.
These results demonstrate that matter-wave interferometry provides a powerful full-distribution function approach for identifying the dynamics of two-dimensional quantum fluids, extending experimental access from the more commonly available two-point correlation analysis methods to higher-order correlations.

The intrinsic access to higher-order correlations of 2D quantum systems and their precise interferometric probe methodology demonstrated in this work has a wide variety of applications, including the understanding of prethermalized states~\cite{AduSmith2013, Murad2022}, phase-ordering dynamics in 2D Bose gases~\cite{comaron_quench_2019}, the sine-Gordon model~\cite{chang2025}, quantum critical points ~\cite{Carleo2013, yu_observing_2024}, dimensional crossover~\cite{Chomaz2015, keepfer2022}, and systems with long-range interactions~\cite{he_exploring_2025}. 
Furthermore, precise full-distribution measurement is a promising probe for the Bose glass phases~\cite{colbois_extreme_2025}.


\section*{Acknowledgements}
We greatly acknowledge Fabian Essler, Vijay Pal Singh, and Ludwig Mathey for insightful discussions.
This work was supported by the EPSRC Grant Reference EP/X024601/1.
A. B. and E. R. thank the EPSRC for doctoral training funding.

\newpage
\section*{Methods}

\section*{Relation of contrast observable to in-situ Bose fields}
The operator corresponding to atomic density after time-of-flight can be approximated as \cite{Polkovnikov2006, Rath2010} 
\begin{align}
    \hat{n}(x,y,z)& \approx n_0(z) \big( \hat{n}_1(x,y)+\hat{n}_2(x,y) \nonumber \\
    &+\hat{A}(x,y)e^{iQz}+\mathrm{h.c.} \big), 
    \label{eq:tof_density_operator}
\end{align}
where $n_0(z)$ is an envelope function, $\hat{n}_{1,2}(x,y)$ are the in-trap atomic density operators corresponding to layers 1 and 2, $\hat{A}(x,y) = \hat{\psi}^\dagger_1(x,y)\hat{\psi}_2(x,y)$ is the interference term, and $Q=m\Delta z/\hbar t_{TOF}$ is the wavevector of matter-wave interference fringes produced by two condensates. 
This approximation is accurate if the cloud does not expand significantly in the $x$-$y$ plane, which is true for our near-uniform system having an in-situ radius of $\SI{22.5}{\micro \meter}$ for the employed time-of flight $t_{\text{TOF}}=\SI{17}{\milli \second}$. 
Since the trapped gases are in the quasi-2D regime, the density profile of a single layer along $z$ after time-of-flight is close to a Gaussian of width (FWHM) $w_{TOF} \approx \SI{75}{\micro \meter} \gg \Delta z$. 
Therefore, we can safely approximate the envelope function $n_0(z)$ with a Gaussian of the same width. 
The form of Eq.~\eqref{eq:tof_density_operator} motivates the use of Fourier transforms along $z$ to extract the eigenvalues of $\hat{A}(x,y)$ from the eigenvalues of $\hat{n}(x,y,z)$. 
For the experimental parameters employed, $n_0(z)$ has a negligible Fourier component at wavevector $Q$, allowing the direct extraction of the eigenvalues of $\hat{A}(x,y)$ by a Fourier transform. The absorption imaging and image postprocessing procedure integrates the atomic density along the $x$-$y$ plane; 
therefore, after appropriate normalisation by the detected number of atoms $N$ and taking an ensemble average, we arrive at, 
\begin{equation}
      \hat{C}=\frac{1}{N} \int_\Omega dx dy~  \hat{\psi}^\dagger_1(x,y)\hat{\psi}_2(x,y). 
\end{equation}
This equation can be generalised straightforwardly to give Eq.~\eqref{eq:C_squared observable}.

\section*{Numerical simulation of equilibrium systems}
\label{app:mc_sims}

An ultracold 2D Bose gas can be described by the Hamiltonian 

\begin{align}\label{eq:2d_bose_gas_quantum_hamiltonian}
    \hat{H} &= \int d\textbf{r} \bigg[\frac{\hbar^2}{2m} \boldsymbol{\nabla} \hat{\psi}^{\dagger}(\textbf{r}) \cdot \boldsymbol{\nabla} \hat{\psi}(\textbf{r}) + \frac{g}{2}\hat{\psi}^{\dagger}(\textbf{r})\hat{\psi}^{\dagger}(\textbf{r})\hat{\psi}(\textbf{r})\hat{\psi}(\textbf{r}) \nonumber \\
    & +V(\textbf{r})\hat{\psi}^{\dagger}(\textbf{r})\hat{\psi}(\textbf{r}) \bigg],
\end{align}

\noindent where $\hat{\psi}(\textbf{r})$ and $\hat{\psi}^{\dagger}(\textbf{r})$ are annihilation and creation operators for bosons, respectively, at the position vector $\textbf{r}$ in the $x$-$y$ plane, $g = \Tilde{g}\hbar^2/m$ is the quasi-2D interaction parameter, and $V(\textbf{r})$ is the external potential \cite{Prokofev2001}.
We use a round box potential with an internal radius of $\SI{22.5}{\micro \meter}$ to simulate the equilibrium system.
To generate \textit{in-situ} samples of the fluctuating Bose field for our system, we used the classical field approximation to replace the field operators in Eq.~(\ref{eq:2d_bose_gas_quantum_hamiltonian}) with complex numbers and their complex conjugates \cite{Blakie_2008}.
We mapped the system onto a $128 \times 128$ square lattice with a lattice constant $\delta l =$ 0.5 \textmu m and used first-order differences to evaluate the derivatives.
The lattice constant was chosen to be smaller than the characteristic length-scales of the system (thermal de Broglie wavelength and healing length) to ensure an accurate representation of the continuum limit. 

We employed the Metropolis-Hastings algorithm \cite{Metropolis_1953, Hastings_1970} to draw random states of a finite-temperature 2D Bose gas from the grand canonical ensemble.
This approach accurately captures both vortex and phonon behaviour and has been previously used to support experimental observations \cite{Sunami2022, Weimer_2015}.
We initialised the complex field with the real and imaginary parts set equal to zero.
At every iteration, we chose a random site and generated proposed updates to the real and imaginary parts of the on-site field by drawing two random numbers from a normal distribution with a mean of zero and a standard deviation of $\sigma$. 
We then calculated the grand-canonical weight $p=\exp{ \left( -\left[ \Delta E - \mu \Delta N \right] / k_BT \right)}$, where $k_B$ is the Boltzmann constant, $\mu$ is the chemical potential, and $\Delta E$, $\Delta N$ are the changes in energy and the number of particles in the system arising from the proposed field update.
The update was accepted according to the standard Metropolis-Hastings procedure. 

To ensure thermal equilibrium and uncorrelated fields, we initially ran the algorithm for 200\,000 updates per site for thermalisation and then began sampling fields, with a further 10\,000 updates per site occurring between each field sample.
We tuned $\sigma$ such that the mean acceptance rate of field updates was near 0.25, which yields optimal convergence \cite{Gelman_1997, Roberts_2001}.

\section*{Experimental procedure: equilibrium}
\label{app:exp_proc_equilibrium}
For the experiments in Secs.~\ref{sec:equilibrium}-\ref{sec:higher},
we form the double-well potential for the dressed atoms using a combination of static and radiofrequency (RF) magnetic fields \cite{Harte2018,Perrin2017}. The static field is a quadrupolar magnetic field with cylindrical symmetry about the vertical axis ($z$), and three RF fields with linear polarisation along $x$ are applied to create a multiple-RF (MRF) trap \cite{Bentine2020}. 
Control over the amplitudes and frequencies of RF components allows us to shape the potential from a single well into a double-well potential, as described in Refs. \cite{Bentine2020,Barker2020jphysb,beregi_2024}. 
In this work, we use the RF frequency combination ($f_1,f_2,f_3$) = (7.02, 7.10, 7.18) MHz to realise a well separation of $\Delta z=\SI{4.9}{\micro \meter}$.
As $\Delta z \gg \ell_0$, we can neglect the Josephson coupling between the two wells.
The RF amplitudes are set such that the trap frequencies along $z$ are $\omega_z / 2\pi = \SI{1.1}{\kilo \hertz}$ in each well, with a double-well barrier height of $E_b/h \sim 4\ \SI{}{\kilo \hertz}$. 
Thanks to the accuracy and stability of modern RF generation apparatus, the fluctuations and drifts of the RF signals are kept well under the 1000 ppm level over at least several weeks, enabling a large number of repeated measurements required for this work~\cite{SunamiThesis}.
After loading the atoms into a single-RF dressed potential and performing evaporative cooling, we transfer the atoms into the MRF-dressed potential adiabatically by slowly introducing the other two RF signals, which can be achieved with negligible heating. 
During the RF amplitude ramps in the single-RF dressed stage, we introduce the optical potential over 3 seconds to realise a near-uniform trap in the $x$-$y$ plane; this potential is created by $\SI{532}{\nano \meter}$ laser light, shaped by a spatial light modulator (digital micromirror device), to produce a box-like trap geometry.
The splitting is performed over $\SI{300}{\ms}$ which is significantly longer than the timescale of the in-plane and axial dynamics. 
We optimise the splitting procedure to ensure that the populations in the two wells are equal by maximising the observed matter-wave interference contrast, as described in Ref.\ \cite{Barker2020}. 
After further equilibrating the gases for $\SI{400}{\ms}$, the MRF-dressed potential and the optical potential are turned off simultaneously, releasing the cloud into TOF expansion to observe the matter-wave interference pattern, as shown in Fig.~\ref{fig:overview} \cite{Sunami2022}.
Finally, as described in the main text, to locally probe the density distribution, we apply a sheet of repumping light before absorption imaging that propagates horizontally (in the $x$ direction) with a thickness $L_y = \SI{8}{\micro \metre}$ and a width much larger than the extent of the cloud of atoms \cite{Barker2020}. 
The thickness was chosen to allow for simultaneous local measurement of phase fluctuations and extraction of interference contrast with a high signal-to-noise ratio. 
We ensure that the repumping light passes through the centre of the cloud by moving the light sheet along the direction parallel to the propagation of imaging light to the position where the total absorption signal is maximum.
We repeat the experiments with repumping light covering the entire cloud to determine phase-space density reported in the main text.
	
\section*{Experimental procedure: quenched systems}	
\label{app:quench_experiment_procedure}
For the results presented in Sec.~\ref{sec:quench}, we begin with an ultracold Bose gas of approximately $9.0 \times 10^4$ $^{87}$Rb atoms in the $F=1$ hyperfine ground state, at a temperature of $T = \SI{60}{\nano \kelvin}$ loaded adiabatically into a cylindrically symmetric, single-well quasi-2D potential, as described in Refs.~\cite{Harte2018,Bentine2020,Luksch2019}.
The trap is created by an MRF-dressed potential \cite{Barker2020,Sunami2022} with three RF components ($f_1,f_2,f_3$) = (7.14, 7.2, 7.26) MHz. 
The static quadrupole field $\bm{B}(\bm{r})=b(x\bm{e}_x+y\bm{e}_y-2z\bm{e}_z)$ has a gradient of $b=\SI{145}{G\, cm}^{-1}$.
Unlike the equilibrium experiments, no optical box-potential was used. 
The single-well potential is characterised by trapping frequencies along the radial and axial directions $\omega_r/2\pi = \SI{13}{\hertz}$ and $\omega_z/2\pi = \SI{1}{\kilo \hertz}$.
These experimental conditions give the dimensionless 2D interaction strength $\tilde{g} = \sqrt{8\pi} a_s/\ell_0=0.076$.
These parameters satisfy the quasi-2D conditions $k_B T \lesssim \hbar \omega_z, \mu$ for the parameters used in this paper: the presence of small excited state populations in the $z$ direction at $\hbar \omega_z \sim k_B T$ results in a reduction of the 2D interaction strength by $\sim 15\%$ \cite{Holzmann2008}; however, this changes the BKT critical temperature by less than 4\,\% \cite{Holzmann2010,Fletcher2015}.
We perform thermometry of the system before the quench (in equilibrium) in the single-well by measuring the expansion of the non-condensed part of the density distribution in a time-of-flight measurement \cite{Sunami2022}.
After holding the gas for 400 ms in the single-well trap \cite{Sunami2022,Hadzibabic2006}, we split the cloud into two daughter clouds by changing the vertical trap geometry from a single- to a double-well potential over 12 ms, as illustrated in Fig.~\ref{fig:quench}a.
The duration of the splitting is chosen to be much shorter than the typical timescale for the radial dynamics $2\pi/\omega_r \sim \SI{80}{\milli \second}$ as well as the typical splitting duration used in the equilibrium experiment. 
After splitting, each well contains $N=4.5\times 10^4$ atoms and has a vertical trap frequency of $\omega_z/2\pi = \SI{1}{\kilo \hertz}$, satisfying the quasi-2D condition. 
We ensure an equal population of the two wells by maximising the contrast of matter-wave interference patterns.
Shortly before the splitting, we adjust the radial trapping frequency to $\omega_r/2\pi = \SI{11}{\hertz}$ by changing $b$ over \SI{10}{\milli \second} so that the radial density profiles before and after are closely matched, thus preventing the quench from exciting the monopole mode (radial breathing).
This radial adjustment, as well as the splitting procedure, is performed over a duration much longer than the characteristic timescale for atomic motion in the vertical direction, $ \tau_{\mathrm{trap}} \sim 2\pi/\omega_z = \SI{1}{\milli \second}$, such that the system remains in the 2D limit (negligible excitation along $z$).
Following the splitting, the static quadrupole field has a field gradient of $b=\SI{94}{G\, cm}^{-1}$.
We confirmed that there is no heating from the splitting procedure itself by measuring the expansion dynamics before and after the splitting of a cloud above the quasicondensation point \cite{Sunami2022} and found the same temperature within uncertainties.
The spatial separation of the minima of the double-well is $\Delta z = \SI{7}{\micro \metre} \gg \ell_0$, which ensures the decoupling of the two clouds for the temperature range and trap parameters used in this work.

\bibliography{refs}

\clearpage
\onecolumngrid
\begin{center}
\textbf{\large Supplementary Information}
\vspace{0.5cm}
\end{center}
\twocolumngrid

\setcounter{table}{0}
\setcounter{figure}{0}
\setcounter{section}{0}
\setcounter{page}{1}
\renewcommand{\thetable}{A\arabic{table}}
\renewcommand{\thefigure}{A\arabic{figure}}
\renewcommand{\theHtable}{Supplement.\thetable}
\renewcommand{\theHfigure}{Supplement.\thefigure}

\section{Contrast extraction and effect of experimental noise}
\label{app:noise}
\begin{figure}[t]
    \includegraphics[width=0.99	\linewidth]{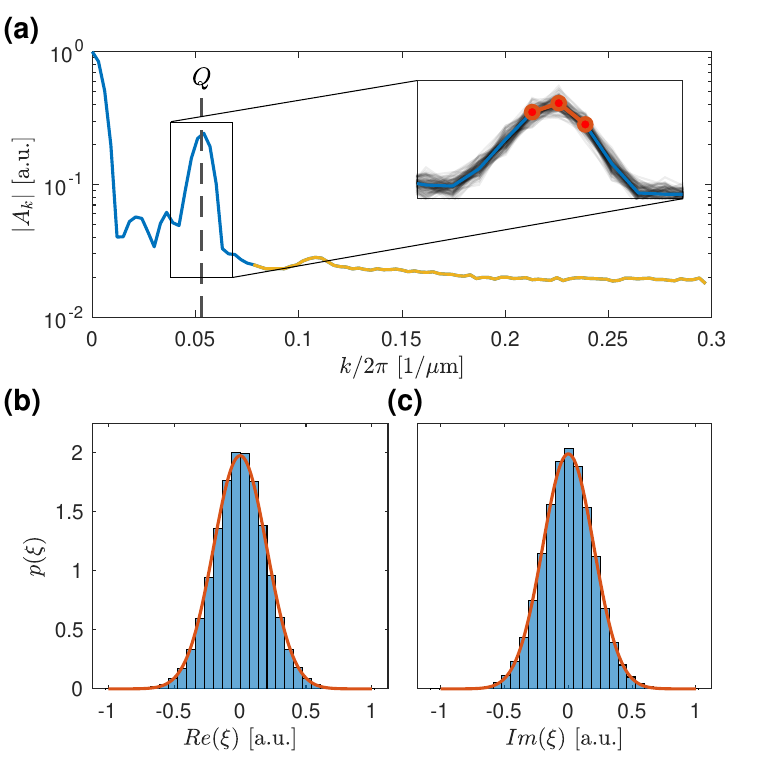}
    \caption{Obtaining noise statistics from images.
    (a), Typical averaged spectrum of interference signal, showing peak at $Q$. Noise statistics are obtained from the part of the spectrum highlighted with yellow. Inset: averaged spectrum in the vicinity of $Q$, plotted on a linear scale. We use the mean of the three highlighted datapoints to calculate the contrast. The shading illustrates realizations of the imaging noise.
    (b),(c), Statistics of the real and imaginary part of contrast noise obtained from the spectrum of experimental images at high wavevectors, with Gaussian fits. }
    \label{fig:sup_contrast_noise}
\end{figure}

The primary source of noise that affects contrast measurements is random noise in the images.
The contrast is extracted as the modulus of the discrete Fourier transform of the integrated column density near the wavevector $Q=m\Delta z/\hbar t_{TOF}$, normalised by the number of atoms detected.
We averaged over $\pm 1$ datapoints around the peak, yielding more accurate values as it averages out fluctuations, as shown in Fig.~\ref{fig:sup_contrast_noise}a.
Using more data points further away from $Q$ is not advantageous, as the signal-to-noise ratio drops.
This results in a small, deterministic decrease in the measured contrast. 
The contrast extraction from a noisy signal is modelled as 
\begin{equation}
    C_{ex} = \vert C_0+\xi_{Re}+i \xi_{Im} \vert, 
    \label{eq:noisy_contrast}
\end{equation}
where $C_0$ is the true contrast, and $\xi_{Re/Im}$ represents the real and imaginary parts of the noise. 
Without loss of generality, we assume $C_0 \in \mathbb{R}$.
We obtain statistics for $\xi_{Re/Im}$ from the Fourier transformed signal at wavevectors $k>1.5Q$, that is very close to a normal distribution, as shown in Fig.~\ref{fig:sup_contrast_noise}b-c, thus it can be accurately modelled by Gaussian white noise.
From Eq.~\ref{eq:noisy_contrast}, it is clear that the noise is not additive; therefore, its effect cannot be simply modelled by the convolution of probability distributions.
This would not be the case if we analysed contrast in the complex domain, but the increased dimensionality of the problem requires significantly more experimental data. 

\begin{figure}[t]
    \includegraphics[width=0.99	\linewidth]{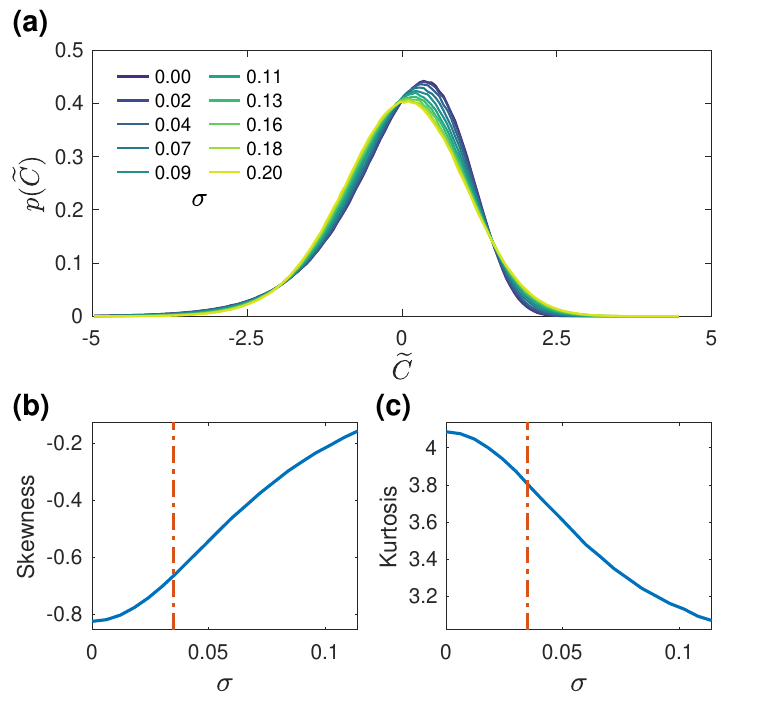}
    \caption{Effect of noise on contrast statistics.
    (a), The theoretical low-temperature universal distribution with various levels of noise relative to the mean contrast.
    (b),(c), Skewness and kurtosis as a function of noise strength affecting the theoretical distribution. The noise strength is relative to the mean contrast. The orange dash-dotted line indicates the experimental noise level at $\mathcal{D}=20.9$. 
    }
    \label{fig:sup_contrast_noise_dist}
\end{figure}

To investigate the systematic effect of imaging noise on the distributions and moments, we drew a large sample from the theoretical low-temperature order parameter distribution.
We adjusted the mean and standard deviation of the generated sample to match the experimentally observed contrast statistics at $\mathcal{D}=20.9$.
The real and imaginary parts of the noise were drawn from a Gaussian distribution with various standard deviations $\sigma$, and the contrast was calculated using Eq.~\ref{eq:noisy_contrast}.
The resulting distributions were normalised to a zero mean and unit variance and compared to the original distribution in Figure~\ref{fig:sup_contrast_noise_dist}a.
Due to the sharper decay of the right tail of the distribution, the most significant effect of the noise is the reduction of negative skewness. 
Furthermore, we numerically calculated the third and fourth central moments of the noisy sample as a function of $\sigma$, as shown in Fig.~\ref{fig:sup_contrast_noise_dist}b-c. 
The procedure for obtaining the corrected distributions in Fig.~\ref{fig:equilibrium_bkt_contrast}a was similar; however, in this case, the original noise-free distribution is obtained from Monte Carlo simulations. 

\section{Statistical analysis of the contrast distributions}
\label{app:statistical_methods}
To analyse and visualise the distributions of observed contrast values, we employ a range of statistical tools, including kernel density estimates (KDEs), empirical cumulative distribution functions (eCDFs), and empirical survival functions (eSFs).
KDEs provide a smooth, continuous estimate of the probability density function from discrete observations and are used here primarily for visualisation.
The confidence intervals KDEs are obtained by bootstrapping.
To calculate KDEs, Gaussian kernels are used throughout, with bandwidths selected according to Scott's rule \cite{scott_kde}.

To compare probability distributions, we utilise eCDFs and their complementary form, eSFs, defined as $\mathrm{eSF} = 1 - \mathrm{eCDF}$. 
These step functions increase by $1/n$ at each of the $n$ observed data points. This method is particularly advantageous, as the calculation of eCDFs does not rely on binning, and they converge to the underlying distribution almost surely.
The confidence intervals for eCDFs and eSFs are obtained with bootstrapping.
Plotting eCDFs and eSFs on logarithmic scales highlights the behaviour of the tails of distributions: specifically, the left tail for eCDFs and the right tail for eSFs.
This approach is used to demonstrate the convergence of the measured distributions to the universal form in Fig.~\ref{fig:equilibrium_bkt_contrast}.

\section{Finite-size scaling analysis of higher-moment observables}
\label{app:finite_size_mc}
\begin{figure}[t]
    \includegraphics[width=0.95	\linewidth]{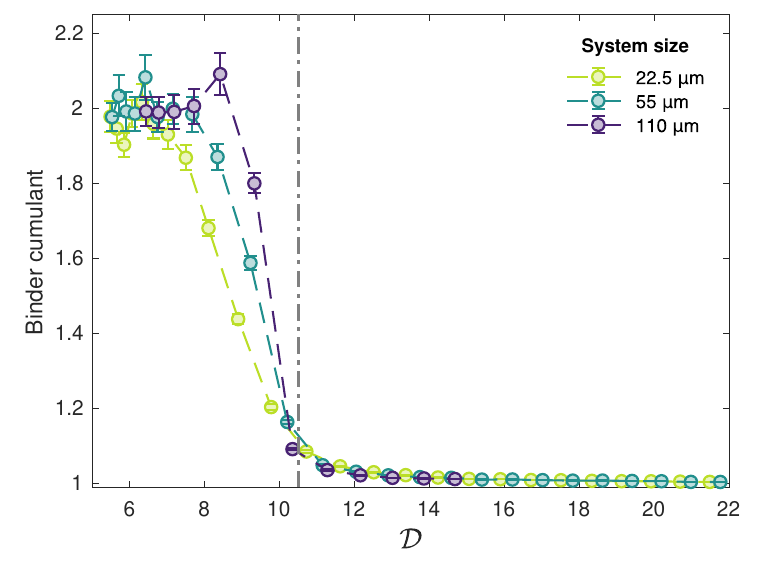}
    \caption{\label{fig:mc_scaling} 
    Finite size scaling analysis of the crossover behaviour of the Binder cumulant, using classical field Monte Carlo simulations. 
    Each datapoint is calculated from 40\,000 realisations of the fluctuating Bose field, and the error bars were obtained via bootstrapping. 
    The grey dash-dotted line indicates the critical point. 
    }
\end{figure}

To analyse the finite-size scaling behaviour of the skewness, kurtosis, and the Binder cumulant, we perform Monte Carlo simulations of 2D Bose gases for various system sizes $r_{sys} = \{ \SI{22.5}{\micro \meter}, \SI{55}{\micro \meter}, \SI{110}{\micro \meter}\}$, the first being the system size for the experiment.
Furthermore, we compare the results with a non-uniform, harmonically trapped system that corresponds to the experimental parameters used in the quench experiment.
We extract the moments in two distinct ways. 
First, for each realisation of the fluctuating Bose field, we calculate the spatially averaged order parameter $A = \int d\mathbf{r} \psi(\mathbf{r})$ within a circular region of radius $r=r_{sys}-\SI{5}{\micro \meter}$, to avoid edge effects. 
Using 40\,000 realisations of $A$, we calculate the Binder cumulant, which we show in Figure~\ref{fig:mc_scaling}. 
The increasing sharpness of the jump in the Binder cumulant allows us to identify precisely the critical phase space density as $\mathcal{D}_c^{(sim)} = 10.5$, indicated by the vertical dash-dotted line. 

\begin{figure*}[t]
    \includegraphics[width=0.95	\linewidth]{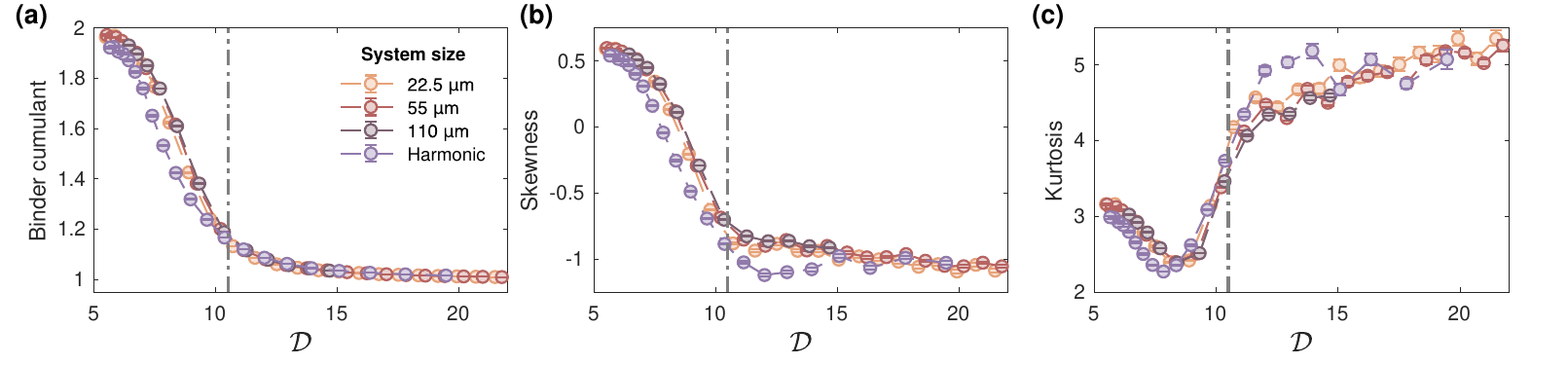}
    \caption{\label{fig:mc_scaling_2} 
    Higher moment observables extracted from simulated interference contrast statistics for various system sizes and for harmonically trapped systems. 
    For all plots, 150\,000 realisations of contrast contributes to each data point. 
     The grey dashed line indicates the critical point, obtained from the finite-size scaling analysis of the Binder cumulant (Figure~\ref{fig:mc_scaling}). 
    }
\end{figure*}

We also model the experiment by evaluating the statistics of interference contrast using Eq.~\eqref{eq:C_squared observable}, with an integration region that matches the experimental selective imaging region, and we calculate the moments from the resulting set of contrast values. 
For this analysis, we randomly select pairs of fields at the same temperature and chemical potential and evaluate the contrast from multiple regions, which provide at least 150\,000 statistically independent contrast values.
Figure~\ref{fig:mc_scaling_2} shows the results for the Binder cumulant, as well as the skewness and kurtosis of the normalised contrast $\widetilde{C}$.
All higher moment observables evaluated from simulated contrast statistics demonstrate good agreement for the range of PSDs for all uniform systems considered.
Non-uniform systems, such as the harmonically trapped one considered here, require an alternative definition for the order parameter \cite{bezett_critical_2009}.
However, as the selective detection region is located near the centre of the system and small compared to the typical system size, the Binder cumulant extracted from the moments of contrast statistics for harmonically trapped systems yields consistent values in the superfluid phase, and the deviations remain small in the crossover region.
Using the previously identified critical point, we find the Binder cumulant's critical value for the interferometric contrast probe to be $ (\mu_4/\mu_2^2)_c^{sim} \approx 1.2$, which is in agreement with previous numerical works in finite-sized systems \cite{davis_microcanonical_2003, keepfer2022}.

\section{Finite-sampling effects on higher-moment observables} 

\begin{figure}[h]
    \includegraphics[width=0.99	\linewidth]{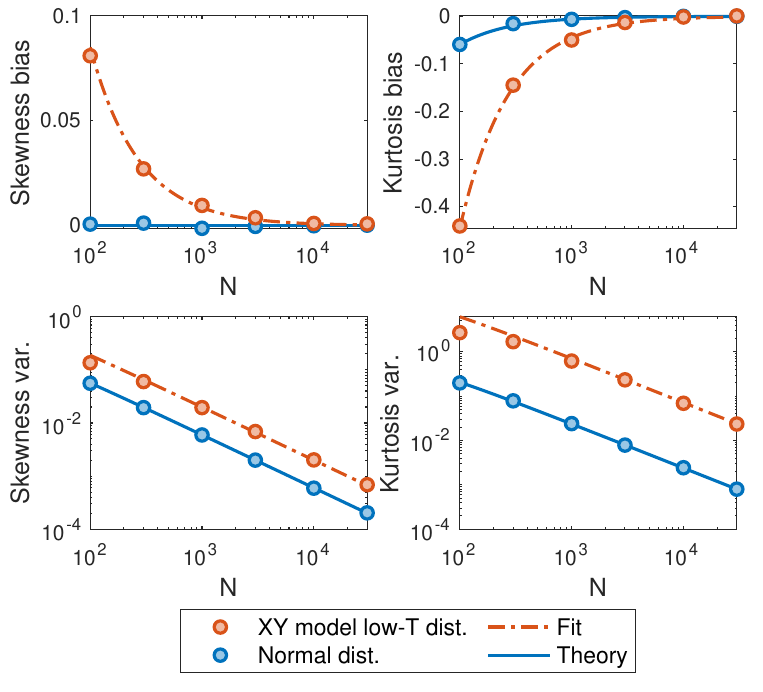}
    \caption{Effect of finite-sampling on higher moment observables, for the low-temperature order parameter distribution of the 2D XY model. 
    }
    \label{fig:sup_finite_sample}
\end{figure}

The characterisation of non-Gaussian distributions from finite-size samples is inherently difficult, as the estimators for skewness and kurtosis can be biased and have a large variance.
To investigate these effects, we generate samples of various sizes from the universal low-temperature distribution of the 2D XY model (Eq. \eqref{eq:bhp_dist}, often called the BHP distribution) using rejection sampling.
We normalise each sample to zero mean, unit standard deviation, and calculate the skewness and kurtosis, similar to the analysis procedure for experimental contrasts.
This procedure is repeated 10\,000 times to obtain accurate statistics on the sampling distribution of the higher moments, and we compute the resulting mean and variance. 
For reference, we repeat the same procedure for samples from a normal distribution. 
We compare the numerical results to the analytical formulae valid for sampling normal distributions \cite{pearson_i_1931}: 
\begin{equation}
    \langle \gamma_1^{(N)} \rangle -\langle \gamma_1^{(\infty)} \rangle = 0,
\end{equation}
\begin{equation}
    \vspace{3pt}
    \langle \gamma_2^{(N)} \rangle -\langle \gamma_2^{(\infty)} \rangle  = -\frac{6}{N+1},
\end{equation}
\begin{equation}
    \langle (\gamma_1^{(N)}- \langle \gamma_1^{(N)}\rangle)^2 \rangle = \frac{6(N-2)}{(N+1)(N+3)},
\end{equation}
\begin{equation}
    \langle (\gamma_2^{(N)}- \langle \gamma_2^{(N)}\rangle)^2 \rangle = \frac{24N(N-2)(N-3)}{(N+1)^2(N+3)(N+5)}.
\end{equation}
We show the results in Figure~\ref{fig:sup_finite_sample}. 
As expected, the reference data sampled from a normal distribution matches the theoretical predictions.
However, the numerical experiments with the BHP distribution yield larger biases and increased variance compared to the normal distribution with zero mean and unit variance. 
We fit the numerical results with $1/N$ models and obtain that for the BHP distribution, 

\begin{equation}
    \langle \gamma_1^{(N)} \rangle-\langle \gamma_1^{(\infty)} \rangle \approx \frac{4.25}{N},
\end{equation}
\begin{equation}
    \langle \gamma_2^{(N)} \rangle-\langle \gamma_2^{(\infty)} \rangle \approx -\frac{45}{N},
\end{equation}
\begin{equation}
    \langle (\gamma_1^{(N)}- \langle \gamma_1^{(N)}\rangle)^2 \rangle \approx \frac{20}{N},
\end{equation}
\begin{equation}
    \langle (\gamma_2^{(N)}- \langle \gamma_2^{(N)}\rangle)^2 \rangle \approx \frac{720}{N}.
\end{equation}
For the typical number of experimental contrast values $N>1000$, this means that for skewness we expect a bias of $8.5\times10^{-3}$ and a standard deviation of $0.14$. 
Similarly, for kurtosis, we expect a bias of -0.045 and a standard deviation of 0.85. 

\section{Real-time renormalisation group equations}

In Eq.~\eqref{eq:real_time_rg_eqns}, we present the real-time RG equations introduced in Ref.~\cite{Sunami2023}, which are based on the equations derived in Refs.~\cite{Mathey2010, Mathey2017} for the dynamical sine-Gordon model, a dual model that describes the BKT transition,
\begin{align}
\frac{dg}{d\ell}&= \left(2-\frac{2}{\tau}\right)g,   \label{eq:rg1} \\
\frac{d\tau}{d\ell}&=\frac{\pi^2 g^2}{\tau}, \label{eq:rg2} 
\end{align}
where $\ell$ is the flow parameter, related to real time by $t=t_0e^\ell$ and $\tau = 4\eta$. 
The non-universal constant of the RG equation in \cite{Mathey2017} is set so that the numerical prefactor for Eq.\ \ref{eq:rg2} is unity. 
These equations describe the dynamics of 2D Bose gases across the critical point, in contrast with the static RG equations used for identifying fixed points that correspond to ordered phases.

\end{document}